# Alterations of brain tissue structural complexity and disorder in Parkinson's disease (PD): Fractal, multifractal, fractal transformation, and disorder strength analyses

Santanu Maity[1], Mousa Alrubayan[1], Mohammad Moshahid Khan[2], and Prabhakar Pradhan[1]

[1]Department of Physics and Astronomy, Mississippi State University, Mississippi State, MS, 39762

[2]Department of Neurology, College of Medicine, University of Tennessee Health Science Center, Memphis, TN 38163

**Abstract**

Parkinson's disease (PD) is marked by progressive neurodegeneration, yet early and subtle structural alterations in brain tissue remain difficult to detect with conventional imaging and analytical methods. Fractal and multifractal frameworks offer a principled way to quantify complex biological architecture, but their diagnostic utility in PD has been largely unexplored. In this study, we investigated the fractal and multifractal characteristics of human brain tissues to identify structural alterations associated with PD. Alongside conventional fractal and multifractal analysis, we employed a recently developed fractal functional distribution method that transforms distributions into a Gaussian form, thereby enhancing quantification. Using this combined approach, we found notable deviations across multiple distribution metrics in PD samples, offering potential for quantitative staging and diagnostic applications. The multifractal analysis revealed threshold-dependent variations in intensity-based measures, which are linked to the sparsity and heterogeneity of neural tissue and suggestive of potential biomarker value. Additionally, we applied inverse participation ratio (IPR) analysis to assess structural disorder, demonstrating that larger IPR pixel sizes correlate with increased structural complexity during disease progression. These complementary analyses outline a multi-layered quantitative profile of PD-related tissue disruption, offering a foundation for earlier, objective assessment of disease-associated microstructural change.

## 1. Introduction

The structural organization of biological tissue is crucial for maintaining physiological functions, spanning nanoscale components to submicron features. As diseases develop, particularly neurodegenerative and neoplastic conditions, tissues often acquire increasing structural irregularity and disorder, reflecting underlying pathological processes [1,2]. These structural alterations usually arise well before clinical symptoms become apparent, motivating increasing interest in quantitative methods capable of detecting early structural disruption and monitoring its evolution as a potential biomarker [3–6]. Bright-field optical transmission microscopy remains a key tool in diagnostic pathology, in which small stained biopsy samples are examined to evaluate morphological abnormalities [7]. However, this conventional approach is more primitive, dependent on human expertise and involves subjective interpretation. As such, it is very susceptible to inter-observer variability and lacks sensitivity to detect the morphological changes that progress with disease [7,8]. In modern pathology, more sophisticated, automated, and quantitative analytical techniques have frequently been developed. These tools are promising and improve traditional diagnostics by offering more statistical analysis of tissue integrity and structure [3,5,9]. High-resolution imaging and computational analysis now make it possible to interrogate nanoscale structural changes, offering a promising avenue for improving early detection, staging, and objective evaluation of diseases ranging from cancer to neurodegenerative disorders [3,10,11].

Parkinson's disease (PD) is a chronic neurodegenerative condition that becomes more common with advancing age, and its global burden continues to grow as life expectancy increases. PD is characterized by the progressive loss of dopaminergic neurons in the substantia nigra accompanied by the accumulation of Lewy bodies containing misfolded α-synuclein. Yet, despite its prevalence, PD is still primarily identified through clinical symptom assessment and conventional imaging and histological examination approaches that often lack the sensitivity and consistency required for early or definitive diagnosis [10,11]. To support the development of more objective diagnostic methods, there is a need for analytical techniques that can capture subtle microstructural changes in affected brain tissue before significant clinical impairment develops. In this work, we performed a detailed analysis of brain tissue samples obtained from both traditional biopsy sections and tissue microarray (TMA) samples to analyze structural alterations in PD. The samples were sectioned at 5 microns thickness, and were selected for scanning to ensure sufficient resolution to probe submicron-scale tissue architectural features [3,10]. To analyze these patterns, we apply fractal-based computational methods, which provide quantitative measures of structural complexity and heterogeneity and have been increasingly explored for their potential to characterize disease-related tissue alterations [12].

Fractal analysis is the concept of self-similarity across different length scales, providing a valuable method for measuring the complex structural irregularities in density variations observed in biological tissues. The complexity of the structure is quantified by fractal dimension ($D_f$), a numerical descriptor that reflects how structural organization changes as disease alters the arrangement and density of cellular components. As the disease progresses, the tissue's structural complexity changes due to rearrangements of its cellular and tissue-level building blocks, as well as the accumulation of additional mass; therefore, the fractal dimension can serve as a numerical measure or quantitative biomarker to monitor the disease state. In addition to exhibiting fractal characteristics, biological tissues frequently display multifractal behavior, arising from the coexistence of multiple scaling patterns. To capture this richer heterogeneity, a multifractal analysis examines the pixel-based intensity distribution using the f(α) vs α spectrum, or Hurst exponent, for detecting more structural complexity. Instead of a single scaling behavior, multifractal analysis captures the spectrum of scaling behaviors to study more heterogeneous and complex structures that scale differently across various regions of tissue. This approach has been particularly useful in detecting subtle transitions and stage-dependent changes from healthy to disease states [12–14].

To further improve our analysis, we applied our established fractal functional transformation, which can generate fractal functional distributions in Gaussian space. This transformation enables more robust statistical interpretation and facilitates more precise differentiation between control and samples at various disease stages [3,12,15]. Complementing these methods, we incorporated the Inverse Participation Ratio (IPR), which serves as a quantitative measure of structural disorder by evaluating the localization characteristics of light within tissue microenvironments [16,17]. Our study presents a multiparametric and image-based quantitative analysis to detect structural changes and disorders in brain tissues affected by PD. By combining fractal, multifractal analysis, fractal functional transformation, and IPR-based analysis, we demonstrate that these techniques have the potential to act as biomarkers for disease detection and staging, offering a promising improvement in clinical pathology [12,13,15,16].

## 2. Methods and Results

### 2.1. Transmission Optical Microscopy of the Brain TMA Samples

#### 2.1.1. Sample Collection

Brain samples are obtained from patients with PD and standard controls through the Biochain Institute Inc. The PD samples from Biochain consist of several cores per slide, one for control and one for diseased tissue. These TMA samples are five μm thick, 1.5 mm in diameter, and mounted on glass slides. These samples are from the hippocampus of postmortem PD patients' brains.

#### 2.1.2. Optical Transmission Microscopy: for automated, fast scanning of the TMA images

We used an automated Olympus BX61 bright-field optical transmission microscope (Tokyo, Japan) to scan and analyze images of biological tissue samples. In previous studies on cancer detection, we demonstrated that changes in the refractive index of biological tissues are linearly related to variations in their mass density. Specifically, for thin tissue samples, the transmission intensity is proportional to the mass density along its width. The relationship between transmission intensity and mass density can be expressed as follows [3]:

$$I_t \propto (n_{tissue+glass} - n_{air}) \propto changes\ in\ tissue\ mass\ density.$$

The variation in mass density, as defined by the above equation, is directly related to changes in the sample's mass density, establishing a link between the sample's mass density variation and the image's transmission intensity. Therefore, by analyzing transmission intensity, we can determine the fractal dimension associated with mass density variation from transmission optical microscopy.

**2.2. Fractal Dimension Analysis of Thin Tissue Samples**

**2.2.1. Fractal Dimension Calculation Using the Box Counting Method**

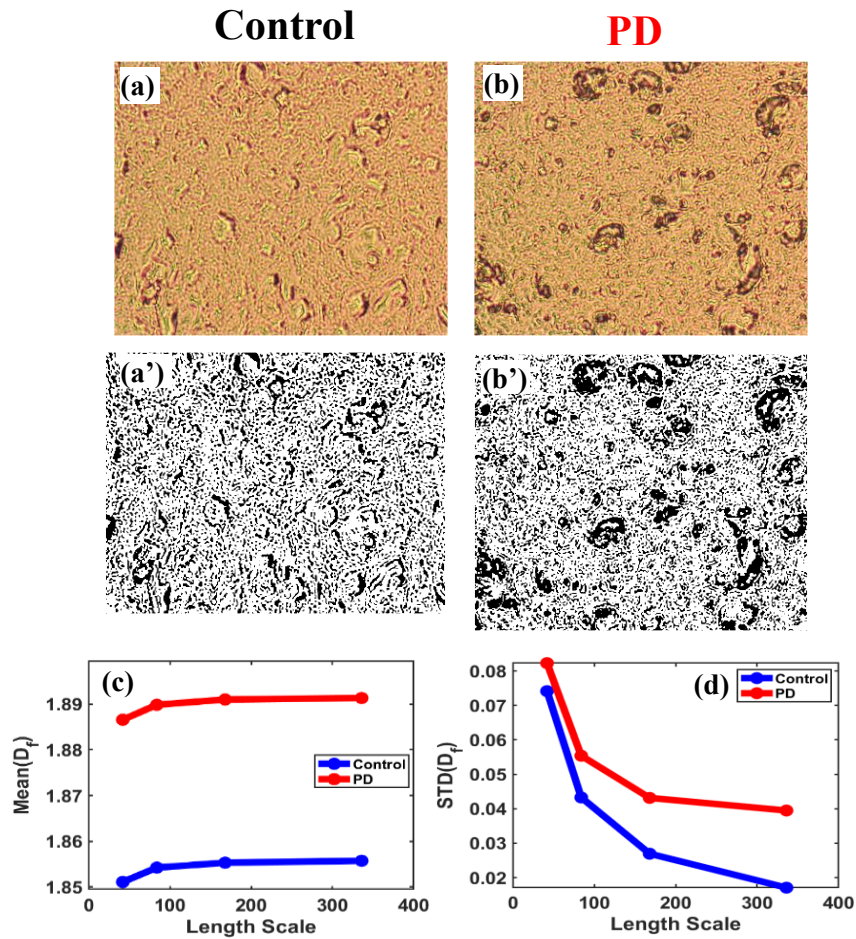

**Figure 1.** (a) and (b) Brightfield images of control and PD brain tissue, respectively, and (a') and (b') are their corresponding binary images. (c) Mean and (d) standard deviation of the local fractal dimension $D_f$ plotted as a function of length scale (region size) for control and Parkinson's disease (PD) brain tissue images.

The box counting method is the standard way to calculate the fractal dimension of biological tissue samples. In this process, grayscale images of tissue microarray (TMA) samples are first converted to 2D binary images, and then the box-counting algorithm is applied to these images.

As described in our previous study, [3] the fractal dimension $D_f$ is calculated using the following equation:

$$D_f = \frac{ln(N(r))}{ln\left(\frac{1}{r}\right)}$$

Where *N(r)* denotes the number of non-empty boxes needed to cover the structure with box size r.

The local fractal dimension is calculated using the box-counting method with varying box sizes and length scales, then averaging the resulting slopes. The region sizes correspond to the varying dimensions (42×42, 84×84, 168×168, and 336×336 pixels) used in the fractal dimension analysis. The consistent trends across regions of various sizes demonstrate the scale-variant nature of the fractal dimension in distinguishing between control and PD brain tissues.

### 2.2.2. Threshold grayscale intensity variation of binary filling in fractal dimension and its effects on fractal parameters

In our previous study, [15] we identified the threshold as a biomarker for stage detection. Several of our earlier studies [3,10,18], a standard threshold value has been suggested for calculating the fractal dimension. These studies describe the procedure for selecting 50% of the grayscale intensities for binary conversion, assigning pixels with intensities above 50% a value of 1 and those below 50% a value of 0. However, here we demonstrated the threshold-dependent behavior of the fractal dimension, with optimal changes at a specific threshold, and identified the threshold region beyond which it can easily distinguish between control and disease. Brain samples, however, show greater sparsity and more variation in gray-scale intensity. This variation occurs during the transformation to binary images when calculating the fractal dimension. Interestingly, a notable trajectory curve reveals the relationship between thresholds and fractal parameters, specifically the mean fractal dimension, Mean ($D_f$), and its standard deviation, STD ($D_f$). At an optimal threshold, the maximum fractal dimension is observed for both the control and PD groups, highlighting the most significant differences between the two groups.

### 2.2.3. Parametric variation: Mean($D_f$) and STD ($D_f$) variations with grayscale threshold intensity percentage.

In Figure 2, we demonstrate the variation in the ensemble-averaged mean fractal dimension, Mean($D_f$), and its standard deviation, STD($D_f$), as functions of the threshold pixel values, evaluated from 10 micrographs each from the control and PD cases. Beyond the threshold of 37.1%, corresponding to a grayscale value of 95 on the grayscale intensity scale (0-255), a significant distinction is observed between the control and PD groups.

In Figure 2(a), we observed an optimal threshold of 58.59%, at which the average fractal dimension shows the most significant difference between control and PD disease tissue. At this threshold, the Mean (Df) values for the control and disease groups are 1.8511 and 1.8866, respectively.

In Figure 2(b), we observed the largest difference in STD ($D_f$) at an optimal threshold of 39.062%, where the fractal dimension standard deviation shows the largest difference between control and PD disease tissue. At this threshold, STD (Df) values for the control and disease groups are 0.0315 and 0.0438, respectively.

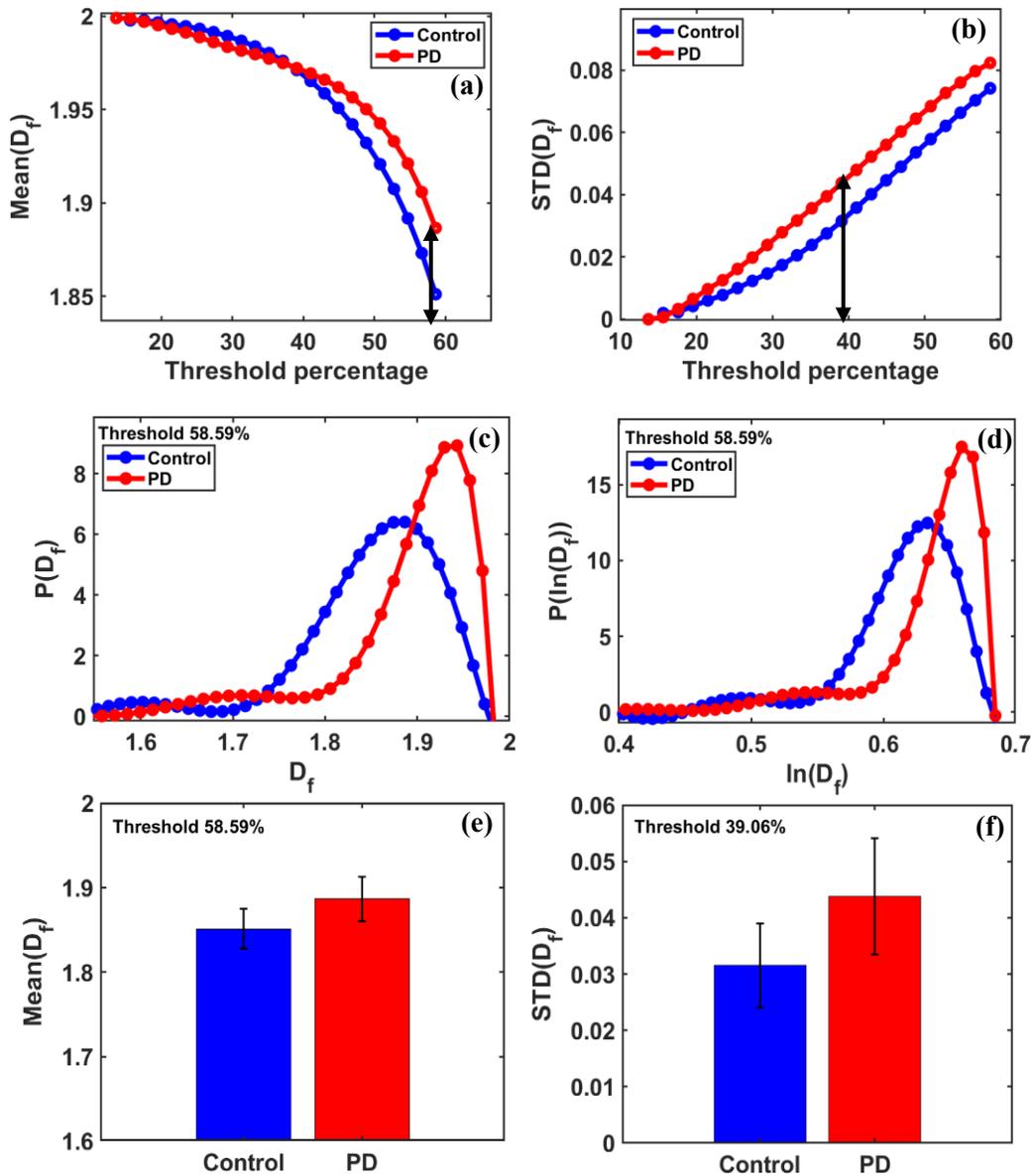

**Figure 2**. (a) Mean fractal dimension ($D_f$) and (b) standard deviation STD ($D_f$) are shown as functions of threshold pixel percentage ranging from 10% to 60% of the grayscale (with default threshold 50%). This illustrates the peak and variation of the peak paths, which differ between the control and PD groups. These differences indicate an optimal threshold at which the maximum difference in $D_f$ values occurs between the two groups. Binary regions of size 42 × 42 pixels used for box-counting fractal dimension analysis with box sizes [2, 3, 6, 7, 14].

(c) The fractal $P(D_f)$ vs. $D_f$ plot shows a slightly extended-tailed distribution. (d) The $P(\ln(D_f))$ vs. $\ln(D_f)$ plot shows a long-tailed non-Gaussian distribution. The most prominent separation between control and PD in the mean values of the $D_f$ curves occurred at a threshold of ~ 58.59%

(e) The bar graphs of the mean (Df) for control and PD groups show that the fractal dimension mean Df value is higher in the PD group than in the control group, with a percentage change of 1.917%. (f) The bar graphs of the standard deviation (STD) for control and PD groups show that the STD (Df) value is in the PD group than in the control group, with a percentage change of 38.742%.

Figure 2(c) shows a slightly extended one-tailed fractal distribution, while Figure 2(d) presents a lognormal transformation, displaying a more extended non-Gaussian distribution. These distributions suggest that the fractal dimension varies systematically with the intensity of the gay scale threshold applied during pre-processing. The optimal fractal dimension occurs at a specific threshold value, which enables the capture of more structural changes for tissue characterization.

**2.2.4. Fractal Analysis of Grayscale Images: Comparison of Optimal Parameters for Control and PD Tissues**

In this study, fractal analysis was performed on grayscale images from both control and diseased tissue samples to examine how fractal dimensions ($D_f$) change with different thresholds and sample sizes. The micrographs are first converted to binary format by applying various threshold values, then divided into regions of various sizes (42×42, 84×84, 168×168, and 336×336 pixels). By applying different grayscale threshold values ranging from 10% to 60% on the threshold intensity scale (0-255), we observed variations in fractal dimension across these binary regions, aiming to understand the influence of the fractal characteristics of the tissues.

To calculate the fractal dimension using the box counting method, different box sizes [2, 3, 6, 7, 14] were applied across the regions to capture fine structural alterations at multiple spatial scales. The analysis was conducted on images of size 672×672 pixels with smaller regions of 42×42 pixels. For instance, in regions such as (1, 2) and (1, 3), which refer to specific coordinates within the brain tissue images, we observed a systematic increasing trend in fractal parameter (Mean ($D_f$)) from the control to the PD brain tissues. We aimed to identify the maximum changes in fractal parameter, such as Mean ($D_f$), between these two groups in smaller, specific regions of tissue structure.

Our analysis showed that localized changes in tissue structure, such as mass localization, occurred in specific regions where the mean fractal dimension ($D_f$) changed significantly. This approach emphasizes the potential of fractal analysis to detect subtle, localized pathological changes within tissue, even when broader structural alterations are not immediately evident.

Additionally, the biological tissues in this study were paraffin-embedded, thereby preserving the overall structure of the samples. Still, it can mask subtle macroscopic pathological changes. However, localized mass formations that may be present can be illustrated and demonstrated through detailed image analysis. Our binary structural representation of tissues revealed discrete regions with structural abnormalities, indicating localized mass accumulation at specific coordinates (e.g., (1,2), (4,3), (5,1), and (5,4)), which showed significant differences and may be of potential pathological relevance.

To further support this analysis, we employed the Inverse Participation Ratio (IPR) technique, which demonstrates that the structural disorder at the nanoscale varies with length scale, providing deeper insights into the heterogeneity of tissue structures in both control and PD brain tissue samples.

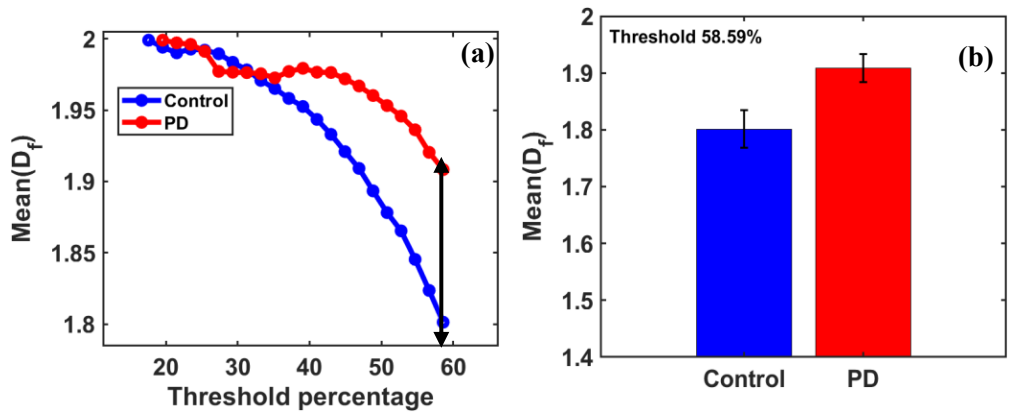

**Figure 3**. (a) For the region at coordinates (1, 2), with a size 42 × 42 pixels, the fractal dimension (Df) was calculated using the box-counting method with box sizes [2, 3, 6, 7, 14]. Mean fractal dimension (Df) as a function of grayscale threshold, pixel threshold percentage ranging from 10% to 60% of the grayscale (with default threshold 50%). This illustrates the peak and variation of the peak paths, which differ between the control and PD groups. These differences indicate an optimal threshold at which the maximum difference in Df values occurs between the two groups.
(b) Bar graphs of the mean (Df) for control and PD groups show that the fractal dimension mean Df value is higher in the PD group than in the control group, with a percentage change of 5.941%.

Figure 3. (a) shows the maximum difference in the mean fractal dimension between control and disease tissues, which occurs at an optimal threshold of 58.59%. At this threshold, the Mean (Df) values for the control and PD tissues are 1.8013 and 1.9083, respectively. Figure 3(b) displays bar graphs of the Mean(Df) for the Control and PD disease groups, showing higher fractal dimension values in the PD group. These preliminary findings aim to identify key regions and optimal parameters for distinguishing control and disease groups based on changes in fractal dimensions. Still, further studies are needed to obtain more conclusive results.

### 2.3. Multifractal Analysis using pixel distributions of the transmission optical microscopy images and f(α) vs. α plot formalism

Brain tissues exhibit a sparser, more multifractal nature. Using the conventional box counting method, fractal analysis enables the study of structural complexity. Due to the inherent sparsity of heterogeneous tissue structure, further multifractal analysis is required.

For multifractal analysis, we employ the probabilistic box-counting method on a 2D brain tissue sample, which is divided into L×L boxes of length scale ε×ε. The probability of pixels present at an *ith* cell can be defined: [12,14,19]

$$P_{\varepsilon,i} = N_\varepsilon(i)/N_{total} \sim \varepsilon^{\alpha_i},$$

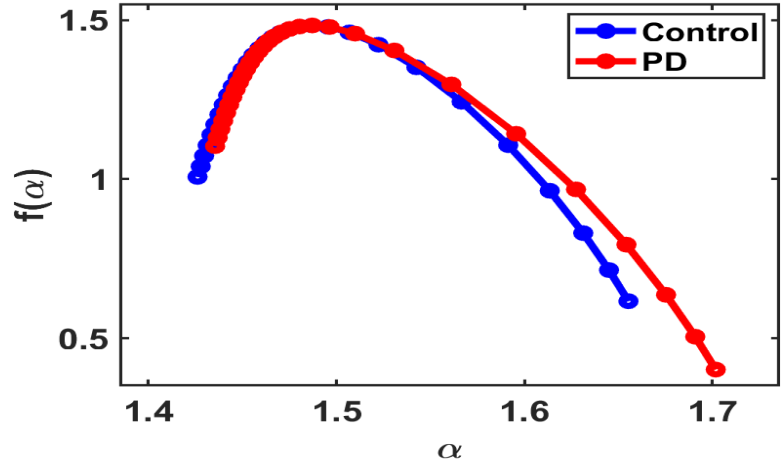

**Figure 4:** Multifractal spectrum f(α) vs. α plot for control and PD groups, each with N=10. This spectrum distribution shows minor variations. (Average Multifractal Spectrum).

$N_\varepsilon(i)$ represents the number of pixels at the ith cell in the length scale ε, and $N_{total}$ represents the total number of pixels.

Following the mathematical procedure gives a multifractal spectrum via then $\mu_{i(Q,\varepsilon)} = P^Q_{i(Q,\varepsilon)} / \sum_{i=1}^{n\varepsilon} P^Q_{i(Q,\varepsilon)}$,

We will plot the spectral function:

$$f(\alpha_Q) = Q \times \alpha_Q - \tau_Q = \sum_{i=1}^{n\varepsilon} (\mu_{i(Q,\varepsilon)} \times \ln(\mu_{i(Q,\varepsilon)})) / \ln \varepsilon.$$

We applied a range for the increased power value Q, scanning from -10 to 10.

Through multifractality analysis, mass density fluctuations scale differently in various regions of the tissue, represented by distinct α values. The multifractal spectrum f(α) describes the distribution of the fractal dimension α, depending on the local scaling exponent α.

Figure 4 shows the average multifractal spectrum, indicating a slightly broader spectrum distribution that highlights the more heterogeneous nature of the PD disease group and its associated structural complexity. In contrast, control tissue samples exhibit more uniform mass distribution.

According to multifractal theory [20] PD disease tissues exhibit a multifractal nature. Due to irregular mass distributions and more non-uniform textures, there are greater structural irregularities than in control tissues. In contrast, control tissues exhibit a uniform distribution of masses, resulting in a uniform texture across all scales.

| Parameter | Control | PD |
|---|---|---|
| $\alpha_{min}$ | 1.4260 | 1.4354 |
| $\alpha_{max}$ | 1.6553 | 1.7018 |
| $\Delta\alpha$ | 0.2293 | 0.2664 |
| $f_{min}$ | 0.616417 | 0.401441 |
| $f_{max}$ | 1.483592 | 1.483592 |
| $\Delta f$ | 0.8672 | 1.0822 |

We extracted the significant spectrum parameters, such as the spectrum's bandwidth ($\Delta\alpha$), height($\Delta f$), maximum singularity exponent ($\alpha_{max}$), and minimum singularity exponent ($\alpha_{min}$), peak singularity strength $f_{max}$, minimum singularity strength $f_{min}$, which collectively represents the strength of multifractality.

Results obtained in the above table support previous findings [15,21], like spectrum bandwidth($\Delta\alpha$), height($\Delta f$), which increased with disease progression, highlighting increased structural disorder and complexity. This method is challenging for quantification because the parameter values change relatively slightly. However, this visual inspection enhances the possibilities of detecting the stages and pathological changes.

**2.4. A new approach: Fractal Functional Transformation**

**2.4.1. Functional transformation of the fractal dimension and its distribution in Gaussian space**

To improve the diagnostic method, we applied a pointwise functional transformation of the fractal dimension to derive quantification parameters that effectively differentiate between brain disease and control.

To implement this approach, we found that the distribution of fractal functional transformation typically follows a long-tailed distribution. We then checked their lognormal distribution, which also shows a closed long tail. We aimed to obtain a Gaussian distribution, as it is statistically simpler to interpret, and its parameters are easier to analyze in terms of average and standard deviation. These quantification parameters are therefore effective for disease detection. The functional transformation equation is given as [12,15].

$$D_{tf} = \frac{D_f}{D_{fmax} - D_f}$$

Where $D_{tf}$ represents the functional transformation of fractal dimension values. For a 2D micrograph, $D_{fmax} = 2$. From our previous studies, we found that the distribution of fractal functional transformation at each point of the fractal dimension is broader and has a long tail, unbounded. After applying their log-normal distribution, we obtained a fractal distribution in Gaussian space, which then becomes a Gaussian. The parameters obtained from the Gaussian distribution, such as Mean and standard deviation of $\ln(D_{tf})$, are diagnostic quantification parameters for brain disease detection. The fractal functional transformation approach is a novel method that shows great potential as a biomarker for detecting brain diseases.

Altogether, the multi-parametric approach for disease detection encompasses fractal analysis, multifractal analysis, threshold-dependent fractal behavior, functional transformation, and inverse participation ratio (IPR) analysis for studying structural disorder and proves to be a cost-effective and highly impactful method for early-stage disease detection.

### 2.4.2. Grayscale threshold variation for binary fractal dimension calculations and its effects on various transform parameters during grayscale threshold scanning of binary fractal filling.

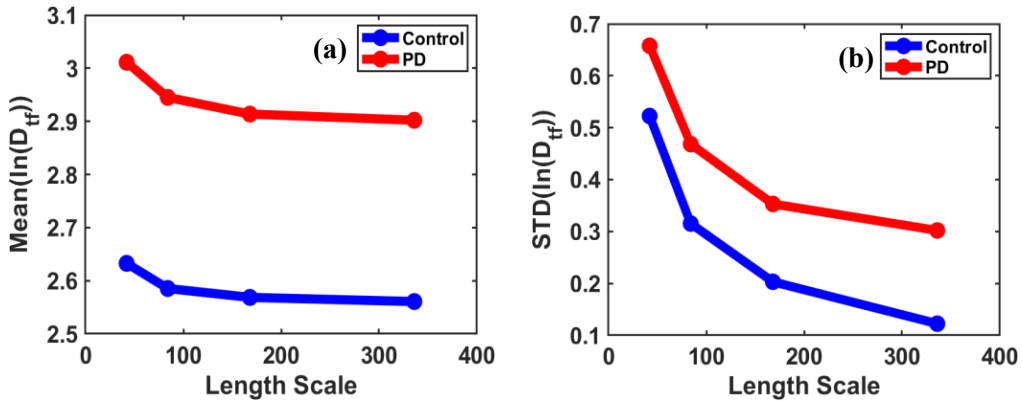

**Figure 5**. (a) Mean(ln(Dtf)) and (b) standard deviation of the functional fractal dimension STD (ln(Dtf)) plotted as a function of length scale (region size) for control and Parkinson's disease (PD) brain tissue images.

Our study found that the fractal dimension varies with threshold during grayscale thresholding (0-255 scale) when scanning 2D binary fractal filling. Grayscale threshold variation produces a unique trajectory that reflects changes in fractal parameters with disease progression. In PD, disease tissue shows a clear distinction from control tissue, with higher optimal values observed above a specific threshold. Similarly, fractal functional parameters also vary with changes in the threshold of the grayscale intensity scale (0-255). This clearly illustrates the relationship between changes in fractal dimension and variations in mass density. Gray-scale threshold variation for binary fractal filling is effectively related to the mass density of pixel values in a 2D fractal. The functional transformation parameters are determined using the box-counting method, applied across smaller regions of the image with varying length scales. Initially, the photos (672×672 pixels) were divided into smaller areas (42×42, 84×84, 168×168, and 336×336 pixels). The consistent patterns observed across these varying length scales demonstrate the scale-variant nature of the functional transformation of fractal dimension, which aids in differentiating control and PD brain tissues.

### 2.4.3. Distribution of $\ln(D_{tf})$ :

Fractal functional parameters, Mean ($\ln(Dtf)$) and STD ($\ln(Dtf)$) values, differ between control and PD disease with varying thresholds. Notably, after a certain threshold, these parameter values increase as the disease progresses. At a specific threshold where these parameters exhibit the maximum difference between the two groups, their distribution follows a normal or Gaussian pattern, facilitating statistical analysis.

These observations of progressive increases in fractal parameters with disease progression demonstrate the diagnostic potential of $\ln(D_{tf})$-based assumptions in distinguishing between control and PD groups, especially at higher threshold levels.

Figure 6 (a) and (b) shows that beyond a threshold range of 31.25% of the maximum threshold value, a clear separation is observed between control and PD disease tissue. In addition, these two classes could be effectively differentiated based on increasing mean values of $\ln(D_{tf})$ and their corresponding standard deviations. Figure 6(a) shows that maximum differences in Mean($\ln(Dtf)$) values between Control and PD diseased tissue occurred at a threshold of 44.92% of the maximum grayscale level. Figure 6(b) shows that maximum differences in STD ($\ln(Dtf)$) values between Control and PD diseased tissue occurred at a threshold of 44.92% of the maximum grayscale level.

Figure 6(c) shows probability distribution plots of $\ln(Dtf)$ for the corresponding maximum Mean ($\ln(D_{tf})$) values. Similarly, Figure 6(d) displays probability distribution plots of $\ln(D_{tf})$ for the corresponding maximum STD ($\ln(D_{tf})$) values. The chi-square test results indicate scores greater than 90% for Gaussian fitting. These results highlight a novel diagnostic technique that utilizes the lognormal distribution of functional transformation parameters, which exhibits a Gaussian distribution.

Figure 6 (e) shows that the Mean($\ln(Dtf)$) values for control and PD are 4.0733 and 4.8333 at the optimal threshold of 44.92%. Figure 6 (f) shows that the STD ($\ln(Dtf)$) values for control and PD are 0.9717 and 1.4131 at the optimal threshold of 44.92%.

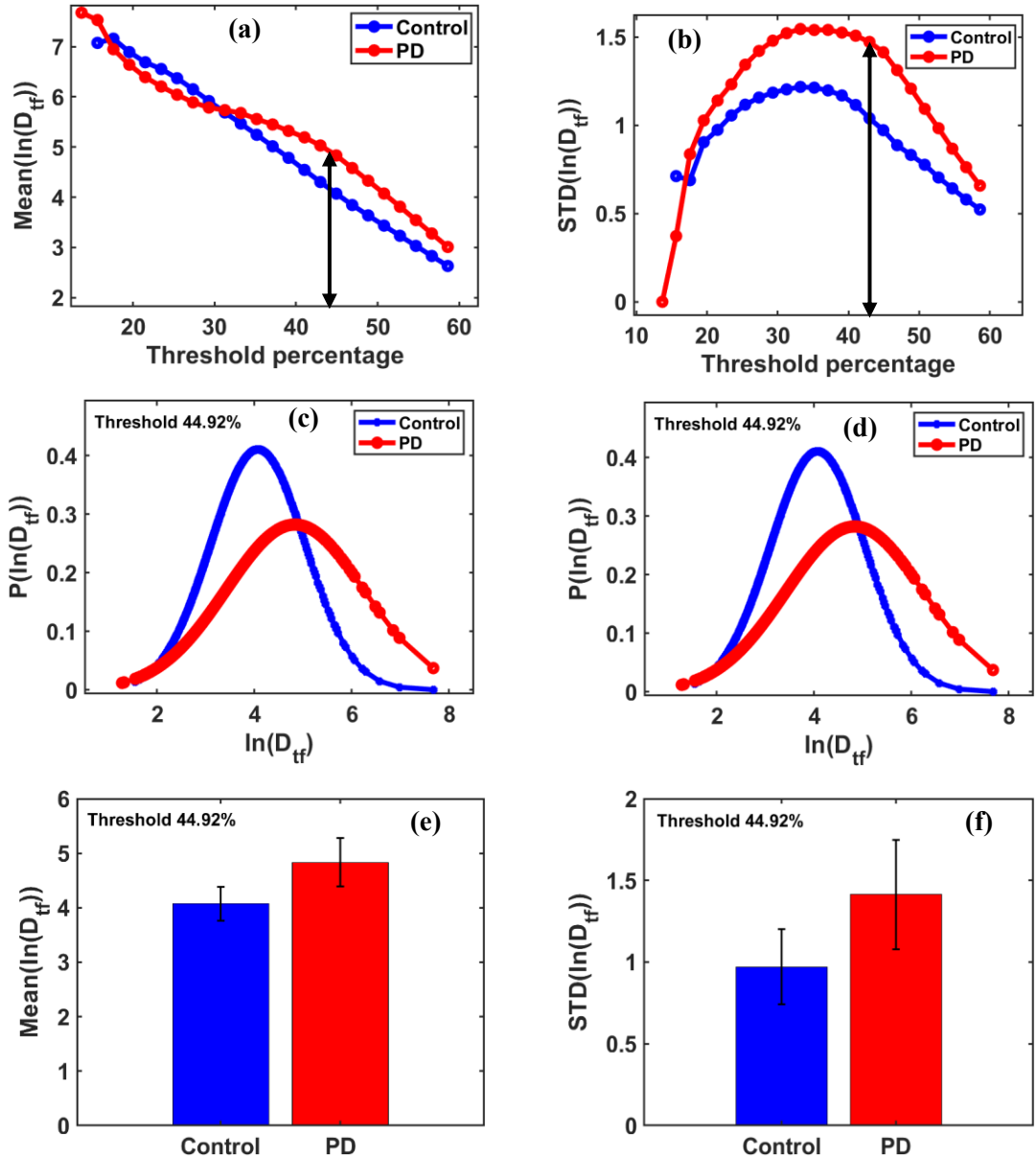

**Figure 6**. (a) Mean ($\ln(D_{tf})$) and (b) STD ($\ln(D_{tf})$) are shown as functions of threshold pixel percentage ranging from 10% to 60% of the grayscale (with default threshold 50%). These threshold-dependent plots exhibit the peak and variation of the peak paths, which differ between the control and PD groups. These differences indicate an optimal threshold value at which the maximum differences in $\ln(Dtf)$ occur between the two groups, suggesting potential diagnostic quantification parameters. Binary regions of size 42 × 42 pixels used for box-counting fractal dimension analysis with box sizes [2, 3, 6, 7, 14].

(c) The plot P($\ln(D_{tf})$) vs ($\ln(D_{tf})$) shows a Gaussian distribution in the best polynomial fits of the Control and PD brain tissues. The chi-square test for Gaussian fitting gives scores > 90%. The most prominent separation between Control and PD in the Mean values of the ln(Dtf) curves occurred at a threshold value of 44.92%.

(d) The plot P($\ln(D_{tf})$) vs $\ln(D_{tf})$ shows a Gaussian distribution in the best polynomial fits of the brain sample for the maximum standard deviation value. The chi-square test for Gaussian fitting gives scores >90%. The most prominent separation between Control and PD in the STD values of the $\ln(D_{tf})$ curves occurred at a threshold value of 44.92%.

(e) The bar graphs of the Mean ($\ln(D_{tf})$) values for the Control and PD groups show that the Mean ($\ln(D_{tf})$) value is higher in the PD group than in the Control group, with a percentage change of 18.656%. (f) The bar graphs of the STD ($\ln(D_{tf})$) of the Control and PD groups show that the STD ($\ln(D_{tf})$) value is higher in the PD group than in the Control group, with a percentage change of 45.419%.

### 2.4.4. Variation of grayscale threshold for binary fractal dimension in different length scale analysis and its impact on functional transformed parameters

We observed that applying fractal analysis to the tissue micrographs for the region at coordinates (1, 2) involved binary conversion of a 42 × 42-pixel image for box-counting fractal dimension analysis with various box sizes [2, 3, 6, 7, 14].

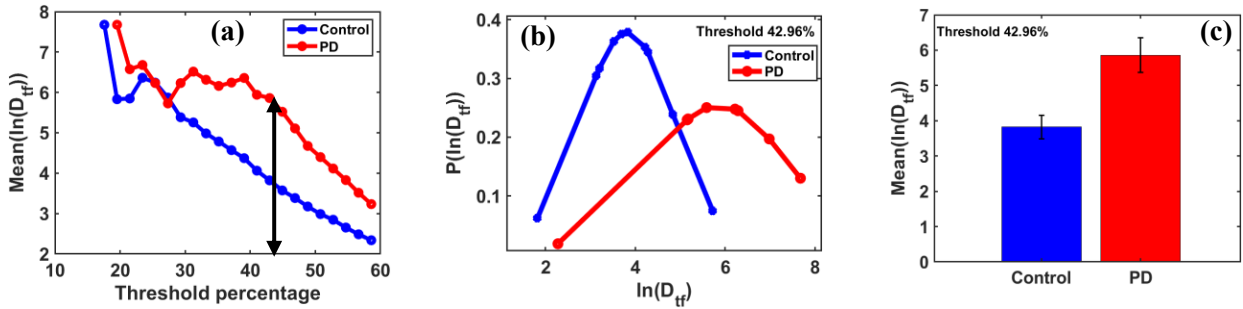

**Figure 7.** (a) For the region at coordinates (1, 2), with a size 42 × 42 pixels, the fractal dimension (Df) was calculated using the box-counting method with box sizes [2, 3, 6, 7, 14]. Mean ($\ln(D_{tf})$) is plotted as a function of the threshold pixel percentage, ranging from 10% to 60% of the grayscale intensity (with a default threshold of 50%). These threshold-dependent plots show the peak and its variation along the peak paths, which differ between the control and PD groups. These differences indicate an optimal threshold at which the maximum differences in Mean (ln(Dtf)) values occur between the two groups, suggesting potential diagnostic quantification parameters.

(b) The plot of P(ln(Dtf)) versus ln(Dtf) shows a Gaussian distribution in the best polynomial fits for both Control and PD brain tissues. The chi-square test for Gaussian fitting yields scores over 90%. The most significant separation between Control and PD in the $\ln(D_{tf})$ curves occurs at a threshold of 42.96%.

(c) The bar graphs of the mean(ln(Dtf)) values for the Control and PD groups indicate that the mean $\ln(D_{tf})$ value is higher in the PD group than in the Control group, with a percentage change of 53.331%.

The mean fractal dimension (Mean(Df)) showed a 5.941% change from control to PD brain tissues.

Additionally, the transformed fractal parameter Mean($\ln(D_{tf})$) also varied with the threshold pixel percentage, ranging from 10% to 60% of the grayscale. These variations in the threshold significantly influenced the fractal transformation parameters, highlighting the threshold-dependent behavior of fractal dimension in distinguishing control and PD brain tissues.

Figure 7 (a) shows that maximum differences in Mean ($\ln(D_{tf})$) values between Control and PD brain tissues occurred at a threshold of 42.96% of the maximum grayscale level.

Figure 7 (b) shows probability distribution plots of ln(Dtf) for the maximum Mean (ln(Dtf)) values in control and PD brain tissues. Figure 7 (c) shows that, at the optimal threshold, the mean (ln(Dtf)) values for the control and PD are 3.8226 and 5.8612, respectively, at a threshold of 42.96%.

**2.5. Structural disorder calculations using transmission intensity using the IPR Technique**

The IPR Technique assesses light localization in the brain tissue samples using 2D micrographs. It is an efficient technique for detecting structural disorder in a system. A low IPR value indicates lower mass density fluctuations in the biological sample's structure. Conversely, a high IPR value suggests that the structure is heterogeneous, with a non-uniform distribution of mass or pixel intensities, indicating higher mass density fluctuations. A high IPR value reflects strong localization, meaning greater structural disorder, which is often observed in diseased tissues due to disrupted mass-density correlations [15,16,22–25].

Changes in mass density $\rho(x, y)$ in a voxel of the cell are directly related to the changes in refractive index, which govern transmission intensity from the tissue sample. It follows the following equation:

$$I((x,y) \propto n(x,y) \propto \rho(x,y)$$

The optical potential can be represented as

$$\varepsilon_i(x,y) = \frac{dn(x,y)}{n_0} \propto \frac{dI(x,y)}{I_0}$$

$I(x, y)$ is the transmission intensity arising from the tissue sample. $n(x, y)$ is denoted as changes in refractive index. This light localization technique conceptually developed from Anderson's tight-binding model. [26,27]

The Hamiltonian of the tight-binding model is represented as

$$H = \sum \varepsilon_i |i><i| + t \sum_{<i,j>} (|i><j| + |i><j|)$$

In the above expression $\varepsilon_i$ is denoted as the potential energy of the optical lattice at the ith lattice site. $|i>$ and $|j>$ are the optical wave functions at the $i^{th}$ and $j^{th}$ lattice sites, ⟨ij⟩ is the nearest neighbor hopping interaction between the $i^{th}$ and $j^{th}$ lattice sites is denoted as t, and t represents the inter-lattice hopping energy parameter.

Mean IPR is defined in the following equation, [17]

$$\langle IPR \rangle_{L \times L} = \frac{1}{N} \sum_{i=1}^{N} \int_0^L \int_0^L E_i^4(x,y) dx$$

In this expression, $E_i$ is the eigenfunction of the Hamiltonian at the ith lattice site. Mean IPR defines the degree of disorder, which is related to the degree of localization.

N is the number of optical potential points in the lattice size L×L. ($N = (L/dx)^2$).

The degree of structural disorder parameter is denoted by $L_d$ and related to changes in refractive index, dn, and correlation length $l_c$. The following equation defines it as $L_d = <dn>l_c$.

After summing up all of these, the following proportionalities can be written,

$$< \langle IPR(L) \rangle_{L \times L} > \propto L_{d-IPR} \sim <dn> \times l_c$$

$$STD(<IPR>L*L) \propto L_{d-IPR} \sim <dn> \times l_c$$

The above relationships provide a quantification method by estimating IPR values, making it easy to detect structural alterations throughout disease progression.

**Results of IPR Analysis: Study on Brain Tissue Samples from Control and PD**

In the IPR analysis, 10 images from each group were randomly selected, and the ensemble-averaged parameters were analyzed. Figure 8. (a1) and (a2) show the brightfield images of control brain tissue samples and corresponding IPR images represented in the IPR color map, which are shown in Figure (a1') and (a2'). Figure 8. (b1) and (b2) show the brightfield images of PD brain tissue samples and corresponding IPR images represented in the IPR color map, which are shown in Figure (b1') and (b2').

Figure 8 (c) and (d) illustrate the nanoscale length-dependent behavior of tissue structure, where increasing nanoscale disorder length results in a more significant distinction between the two groups.

These differences are easily identified by estimating the Mean (IPR) and standard deviation (STD) values for the control and PD disease groups. Figure 8. (c) shows that at higher pixel sizes, Mean IPR has the maximum changes between the control and PD disease groups. Additionally, increasing the pixel size in the results increases IPR values associated with greater structural disorder. Figure 8. (d) At higher pixel sizes, the STD of IPR exhibits the maximum changes between the control and PD disease groups. Additionally, increasing the pixel size in the results increases IPR values associated with greater structural disorder.

Figure 8 (c') displays the bar graphs of the Mean IPR for the Control and PD brain tissues—the Mean IPR value changes by 22.38% from the Control to the PD brain tissues. Figure 8 (d') shows the bar graphs of the STD (IPR) for the Control and PD brain tissues. The standard deviation (STD) of IPR values changes by 71.94% from the Control to the PD brain tissues. This variability in IPR values is high, indicating high structural disorder in PD groups at the nanoscale disorder level. This quantification method is cost-effective, more accurate, and valuable for diagnosing disease at its earliest stages.

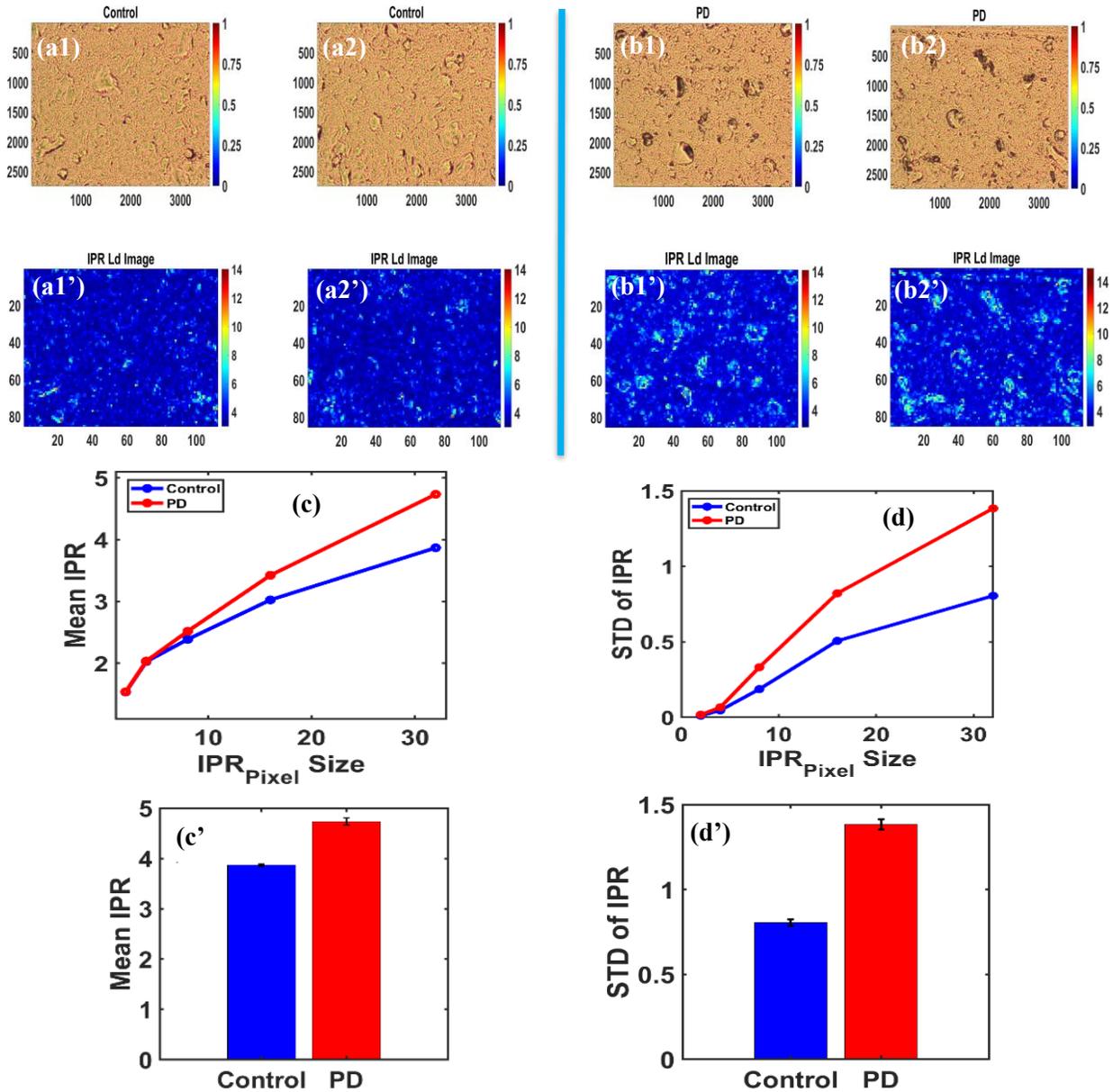

**Figure 8**. (a1) and (a2) are the brightfield images of control brain tissues. Whereas (a1') and (a2') are the respective IPR colormap images of control brain tissues. (b1) and (b2) are the brightfield images of Parkinson's disease tissues. Whereas (b1') and (b2') are the respective IPR colormap images of Parkinson's disease tissues. Changes in Mean IPR values related to structural alterations in tissue are represented by Mean(IPR) ~ $L_{d\text{-}IPR}$ color map. IPR Pixel size 32 is used for this analysis.

(c) shows the Mean IPR as a function of IPR pixel size for Control and PD groups, showing percentage changes of 5.51%, 13.16%, and 22.38% for matrix sizes of 8, 16, and 32, respectively.

(d) exhibits the standard deviation (STD) of IPR as a function of IPR pixel size for control and PD groups, with percentage changes of 66.34%, 44.07%, 78.13%, 62.26%, and 71.94% for matrix sizes of 2, 4, 8, 16, and 32, respectively.

(c') and (d') show the bar graphs of the Mean IPR and the Standard deviation (STD) of IPR for control and PD brain tissues. The Mean IPR value is 22.38% higher in PD brain tissues compared to control tissues. The STD of IPR is 71.94% higher in PD brain tissues than in control tissues.

**3. Discussion and Conclusions:**

In this study, we employed a multi-parametric, image-based analytical framework to examine structural alterations and disorders in brain tissues affected by PD. By integrating fractal analysis, multifractal characterization, fractal functional transformation, and IPR-based light localization methods into bright-field optical images of biopsy and TMA samples, we aimed to develop quantitative metrics capable of distinguishing PD from control tissue and to explore their potential as biomarkers for disease diagnosis and staging. Together, these tools allowed us to probe nanoscale structural heterogeneity that is not readily captured through conventional histopathological or visual inspection approaches. Our findings are consistent with and extend a growing body of literature demonstrating that fractal and related complexity measures are sensitive to pathological tissue changes in both neurodegenerative disease and cancer [6,28–31].

Several studies have shown that fractal dimension (FD) is a useful descriptor of brain structural complexity in neurodegeneration [2,28]. Recent reports have emphasized FD as a promising imaging biomarker across multiple modalities, highlighting its ability to capture subtle structural irregularities that do not appear in conventional volumetric measurements [28,29]. We have previously reported measurable FD changes using transmission optical microscopy of thin sections derived from Alzheimer's and Parkinson's disease postmortem brains [10]. Our work builds on these studies by demonstrating that FD derived from bright-field images of PD brain tissue is not only altered but also exhibits systematic, threshold-dependent behavior, which itself carries discriminative information. We presented the results of PD tissue samples to characterize their properties using our newly developed multi-parametric approach, including fractal, multifractal, fractal functional transformation, and Inverse Participation Ratio analysis. This method involves several quantification parameters derived from the study for disease diagnosis. Most importantly, for PD TMA samples, a direct connection is shown between variation in mass density and changes in refractive index of the sample, which governs the transmission intensity. This mass density variation in a 2D binary fractal is related to its fractal dimension $D_f$.

Exact fractal dimension calculations require the determination of a threshold in grayscale. Interestingly, the fractal dimension emerges as a function of grayscale threshold variation. For PD results, the distributions of multifractal, fractal, and functional analyses of the fractal dimension exhibit well-separated distributions with specific quantitative values, and an optimal value corresponds to the best threshold choices. Our results indicate that the threshold can be considered a potential biomarker for disease detection, and fractal and multifractal parameters, such as Mean ($D_f$) and STD ($D_f$), can serve as quantification parameters to assess disease stages. Additionally, introducing a novel approach, the functional transformation of fractal dimensions follows a log-normal distribution, which offers a new direction for our analysis. Applying the logarithmic transformation produces a normal or Gaussian distribution. The distribution parameters, such as the mean and standard deviation, increase with disease progression. These quantification parameters, Mean ($\ln(D_{tf})$) and STD ($\ln(D_{tf})$), are easily interpreted in the quantitative analysis of brain disease.

To support our diagnostic approach, we utilize our well-known, recently developed light localization method, the Inverse Participation Ratio (IPR), which enhances the possibility of an accurate diagnosis. This method detected changes in structural disorder at the nanoscale level with disease progression. IPR values are linearly related to structural disorder strength, with IPR proportional to $L_d$ and IPR~dn×lc. IPR values increased by increasing length scales or sample size to a certain point; in effect, structural disorder also increased. We have previously shown that IPR applied to TEM images can quantify nanoscale refractive-index fluctuations associated with early carcinogenesis; increases in average IPR were interpreted as elevated nanoscale disorder and were proposed as some of the earliest structural signatures of malignant transformation [22,32]. More recent work has extended IPR-based methods to optical and confocal imaging to probe DNA-specific or molecularly targeted structural alterations [24,33,34]. Our study translates this approach to PD brain tissue using bright-field microscopy, showing that IPR-derived parameters increase with length scale and sample size up to a saturation regime, consistent with progressively stronger structural disorder in PD. By demonstrating that IPR metrics (mean and standard deviation) differ between PD and control tissue, our results support the notion that nanoscale disorder is not unique to cancer but also characterizes neurodegenerative pathology. This is in line with the broader hypothesis that many diseases share an early signature of increased nanoscale refractive index fluctuations, reflecting disrupted cellular and subcellular organization. IPR thus complements fractal and multifractal measures: while FD and multifractal spectra emphasize geometric and intensity-based scaling properties at micro- and mesoscales, IPR directly probes the strength of refractive-index disorder at shorter length scales.

In conclusion, our study supports the feasibility and utility of a multi-parametric, physics-inspired framework for quantifying PD-related structural disorder in brain tissue using standard histological images. By leveraging fractal, multifractal, fractal functional transformation, and IPR-based analyses, we demonstrate that complex structural alterations in PD can be captured through a set of interpretable numerical parameters with strong conceptual ties to previously published work in neurodegeneration and cancer. These methods hold promise as candidate biomarkers for PD detection and staging and may contribute to more objective, early, and quantitative assessments in clinical pathology.

———————————

**Acknowledgments:** We thank the Mississippi State and UTHSC imaging facilities for imaging.
**Funding:** This work was partially supported by the National Institutes of Health (NIH) under grant R21 CA260147 to PP ; and MMK's work was supported by NIH grant number R21NS128519.

## References:

[1] S. S. Cross, FRACTALS IN PATHOLOGY, J. Pathol. **182**, 1 (1997).

[2] E. T. Ziukelis, E. Mak, M.-E. Dounavi, L. Su, and J. T O'Brien, Fractal dimension of the brain in neurodegenerative disease and dementia: A systematic review, Ageing Research Reviews **79**, 101651 (2022).

[3] L. Elkington, P. Adhikari, and P. Pradhan, Fractal Dimension Analysis to Detect the Progress of Cancer Using Transmission Optical Microscopy, Biophysica **2**, 1 (2022).

[4] R. Lopes and N. Betrouni, Fractal and multifractal analysis: A review, Medical Image Analysis **13**, 634 (2009).

[5] A. H. Beck, A. R. Sangoi, S. Leung, R. J. Marinelli, T. O. Nielsen, M. J. Van De Vijver, R. B. West, M. Van De Rijn, and D. Koller, Systematic Analysis of Breast Cancer Morphology Uncovers Stromal Features Associated with Survival, Sci. Transl. Med. **3**, (2011).

[6] L. G. Da Silva, W. R. S. Da Silva Monteiro, T. M. De Aguiar Moreira, M. A. E. Rabelo, E. A. C. P. De Assis, and G. T. De Souza, Fractal dimension analysis as an easy computational approach to improve breast cancer histopathological diagnosis, Appl. Microsc. **51**, 6 (2021).

[7] T. J. Evans and P. A. Riley, Principles of microscopy, culture and serology-based diagnostics, Medicine **49**, 648 (2021).

[8] J. G. Elmore et al., Diagnostic Concordance Among Pathologists Interpreting Breast Biopsy Specimens, JAMA **313**, 1122 (2015).

[9] M.-C. Weber et al., Fractal analysis of extracellular matrix for observer-independent quantification of intestinal fibrosis in Crohn's disease, Sci Rep **14**, 3988 (2024).

[10] I. Apachigawo, D. Solanki, R. Tate, H. Singh, M. M. Khan, and P. Pradhan, Fractal Dimension Analyses to Detect Alzheimer's and Parkinson's Diseases Using Their Thin Brain Tissue Samples via Transmission Optical Microscopy, Biophysica **3**, 569 (2023).

[11] R. B. Postuma et al., MDS clinical diagnostic criteria for Parkinson's disease: MDS-PD Clinical Diagnostic Criteria, Mov Disord. **30**, 1591 (2015).

[12] S. Maity, M. Alrubayan, I. Apachigwao, D. Solanki, and P. Pradhan, *Optical Probing of Fractal and Multifractal Connection to Structural Disorder in Weakly Optical Disordered Media: Application to Cancer Detection*.

[13] A. Karperien, H. Ahammer, and H. F. Jelinek, Quantitating the subtleties of microglial morphology with fractal analysis, Front. Cell. Neurosci. **7**, (2013).

[14] H. F. Jelinek, N. T. Milošević, A. Karperien, and B. Krstonošić, *Box-Counting and Multifractal Analysis in Neuronal and Glial Classification*, in *Advances in Intelligent Control Systems and Computer Science*, edited by L. Dumitrache, Vol. 187 (Springer Berlin Heidelberg, Berlin, Heidelberg, 2013), pp. 177–189.

[15] S. Maity, M. Alrubayan, and P. Pradhan, *Decoding Breast Cancer in X-Ray Mammograms: A Multi-Parameter Approach Using Fractals, Multifractals, and Structural Disorder Analysis*.

[16] P. Pradhan, D. Damania, H. M. Joshi, V. Turzhitsky, H. Subramanian, H. K. Roy, A. Taflove, V. P. Dravid, and V. Backman, Quantification of nanoscale density fluctuations using electron microscopy: Light-localization properties of biological cells, Applied Physics Letters **97**, 243704 (2010).

[17] P. Pradhan and S. Sridhar, Correlations due to Localization in Quantum Eigenfunctions of Disordered Microwave Cavities, Phys. Rev. Lett. **85**, 2360 (2000).

[18] S. Bhandari, S. Choudannavar, E. R. Avery, P. Sahay, and P. Pradhan, Detection of colon cancer stages via fractal dimension analysis of optical transmission imaging of tissue microarrays (TMA), Biomed. Phys. Eng. Express **4**, 6 (2018).

[19] A. Chhabra and R. V. Jensen, Direct determination of the f(α) singularity spectrum, Phys. Rev. Lett. **62**, 1327 (1989).

[20] C. J. Morales and E. D. Kolaczyk, Wavelet-Based Multifractal Analysis of Human Balance, Annals of Biomedical Engineering **30**, 588 (2002).

[21] A. J. Joseph and P. N. Pournami, Multifractal theory based breast tissue characterization for early detection of breast cancer, Chaos, Solitons & Fractals **152**, 111301 (2021).

[22] P. Pradhan, D. Damania, H. M. Joshi, V. Turzhitsky, H. Subramanian, H. K. Roy, A. Taflove, V. P. Dravid, and V. Backman, Quantification of nanoscale density fluctuations by electron microscopy: probing cellular alterations in early carcinogenesis, Phys. Biol. **8**, 026012 (2011).

[23] P. Sahay, A. Ganju, H. M. Almabadi, H. M. Ghimire, M. M. Yallapu, O. Skalli, M. Jaggi, S. C. Chauhan, and P. Pradhan, Quantification of photonic localization properties of targeted nuclear mass density variations: Application in cancer-stage detection, Journal of Biophotonics **11**, e201700257 (2018).

[24] P. Adhikari, P. K. Shukla, F. Alharthi, R. Rao, and P. Pradhan, Photonic technique to study the effects of probiotics on chronic alcoholic brain cells by quantifying their molecular specific structural alterations via confocal imaging, Journal of Biophotonics **15**, e202100247 (2022).

[25] P. Sahay, H. M. Almabadi, H. M. Ghimire, O. Skalli, and P. Pradhan, Light localization properties of weakly disordered optical media using confocal microscopy: application to cancer detection, Opt. Express **25**, 15428 (2017).

[26] W. M. C. Foulkes and R. Haydock, Tight-binding models and density-functional theory, Phys. Rev. B **39**, 12520 (1989).

[27] J. Fröhlich, F. Martinelli, E. Scoppola, and T. Spencer, Constructive proof of localization in the Anderson tight binding model, Commun. Math. Phys. **101**, 21 (1985).

[28] D. Pirici, L. Mogoanta, D. A. Ion, and S. Kumar-Singh, Fractal Analysis in Neurodegenerative Diseases, Adv Neurobiol **36**, 365 (2024).

[29] L. Díaz Beltrán, C. R. Madan, C. Finke, S. Krohn, A. Di Ieva, and F. J. Esteban, Fractal Dimension Analysis in Neurological Disorders: An Overview, Adv Neurobiol **36**, 313 (2024).

[30] E. Roura et al., Cortical fractal dimension predicts disability worsening in Multiple Sclerosis patients, Neuroimage Clin **30**, 102653 (2021).

[31] N. A. Davies et al., Fractal dimension (df) as a new structural biomarker of clot microstructure in different stages of lung cancer, Thromb Haemost **114**, 1251 (2015).

[32] P. Adhikari, M. Hasan, V. Sridhar, D. Roy, and P. Pradhan, Studying nanoscale structural alterations in cancer cells to evaluate ovarian cancer drug treatment, using transmission electron microscopy imaging, Phys Biol **17**, 036005 (2020).

[33] F. Alharthi, I. Apachigawo, D. Solanki, S. Khan, H. Singh, M. M. Khan, and P. Pradhan, Dual Photonics Probing of Nano- to Submicron-Scale Structural Alterations in Human Brain Tissues/Cells and Chromatin/DNA with the Progression of Alzheimer's Disease, Int J Mol Sci **25**, 12211 (2024).

[34] F. Alharthi, D. Solanki, I. Apachigawo, S. Maity, J. Xiao, M. M. Khan, and P. Pradhan, Optical Detection of the Spatial Structural Alteration in the Human Brain Tissues/Cells and DNA/Chromatin due to Parkinson's Disease, J Biophotonics e70131 (2025).